\documentclass[twocolumn,aps,prl,preprintnumbers,superscriptaddress,reprint,floatfix]{revtex4-1}
\usepackage{bbm}
\usepackage{bm}
\usepackage{amsmath}
\usepackage{amssymb}
\usepackage{empheq}
\usepackage{graphicx}
\usepackage{mathrsfs}
\usepackage{multirow}
\usepackage{amsfonts}
\usepackage{amsthm}
\usepackage{color}
\usepackage{bigints}
\usepackage{txfonts}
\usepackage{hyperref}
\hypersetup{
     unicode=false,          
     pdftoolbar=true,        
     pdfmenubar=true,        
     pdffitwindow=false,     
     pdfstartview={FitH},    
     pdftitle={My title},    
     pdfauthor={Author},     
     pdfsubject={Subject},   
     pdfcreator={Creator},   
     pdfproducer={Producer}, 
     pdfkeywords={keyword1} {key2} {key3}, 
     pdfnewwindow=true,      
     colorlinks=false,       
     linkcolor=red,          
     citecolor=green,        
     filecolor=magenta,      
     urlcolor=cyan           
}

\providecommand{\U}[1]{\protect\rule{.1in}{.1in}}
\makeatletter
\@ifundefined{textcolor}{}
{
\definecolor{BLACK}{gray}{0}
\definecolor{WHITE}{gray}{1}
\definecolor{RED}{rgb}{1,0,0}
\definecolor{GREEN}{rgb}{0,1,0}
\definecolor{BLUE}{rgb}{0,0,1}
\definecolor{CYAN}{cmyk}{1,0,0,0}
\definecolor{MAGENTA}{cmyk}{0,1,0,0}
\definecolor{YELLOW}{cmyk}{0,0,1,0}
}
\makeatother

\begin{document}
\title{Disorder-Induced Quantum Phase Transitions in Three-Dimensional Second-Order Topological Insulators}
\author{C. Wang}
\email[Corresponding author: ]{physcwang@tju.edu.cn}
\affiliation{Center for Joint Quantum Studies and Department of 
Physics, School of Science, Tianjin University, Tianjin 300350, China}
\author{X. R. Wang}
\email[Corresponding author: ]{phxwan@ust.hk}
\affiliation{Physics Department, The Hong Kong University of Science 
and Technology (HKUST), Clear Water Bay, Kowloon, Hong Kong}
\affiliation{HKUST Shenzhen Research Institute, Shenzhen 518057, China}
\date{\today}

\begin{abstract}

Disorder effects on three-dimensional second-order topological insulators (3DSOTIs) are investigated numerically 
and analytically. The study is based on a tight-binding Hamiltonian for non-interacting electrons on a cubic 
lattice with a reflection symmetry that supports a 3DSOTI in the absence of disorder. Interestingly, unlike the 
disorder effects on a topological trivial system that can only be either a diffusive metal (DM) or an Anderson 
insulator (AI), disorders can sequentially induce four phases of 3DSOTIs, three-dimensional first-order topological
insulators (3DFOTIs), DMs and AIs. At a weak disorder when the on-site random potential of strength $W$ is below a 
low critical value $W_{c1}$ at which the gap of surface states closes while the bulk sates are still gapped, the 
system is a disordered 3DSOTI characterized by a constant density of states and a quantized integer conductance of 
$e^2/h$ through its chiral hinge states. The gap of the bulk states closes at a higher critical disorder $W_{c2}$, 
and the system is a disordered 3DFOTI in a lower intermediate disorder between $W_{c1}$ and $W_{c2}$ in which 
electron conduction is through the topological surface states. The system becomes a DM in a higher intermediate 
disorder between $W_{c2}$ and $W_{c3}$ above which the states at the Fermi level are localized. It undergoes a 
normal three-dimension metal-to-insulator transition at $W_{c3}$ and becomes the conventional AI for $W>W_{c3}$. 
The self-consistent Born approximation allows one to see how the density of bulk states and the Dirac mass are 
modified by the on-site disorders.
\end{abstract}

\maketitle

\section{Introduction}

Topological states of matter have attracted much attention in condensed matter physics in recent years because 
of their exotic properties such as the topologically protected surface and edge states. These states can 
exist in both topological insulators \cite{kane1,bernevig1,konig1,hasan1,qi1} and Weyl semimetals \cite{wan}. 
Their existences are guaranteed by the bulk-boundary correspondence rooted in the Stokes-Cartan theorem. 
Fermions or Bosons and classical or quantum particles, such as electrons \cite{hasan1,qi1}, phonons 
\cite{susstrunk1,zyliu1,garcia1,xni1,hfan1}, photons \cite{khanikaev1,Lu1,hassan1}, and magnons 
\cite{xiansi1,xiansi2,suying1,suying2}, can have topological states. According to the bulk-boundary 
correspondence, a three-dimensional (3D) insulator with band inversion \cite{qi1} has topologically non-trivial 
two-dimensional surface states. This insulator is a 3D first-order topological insulator (3DFOTI). 
When the surface states of a 3DFOTI are gapped, the intersection of two surfaces of different topological 
classes, i.e. a hinge, has topologically non-trivial chiral hinge states, leading to a so-called 3D second-order 
topological insulator (3DSOTI). The well-accepted paradigm is that a $d$ dimensional material can be a 
$(d-n)$th-order topological insulator with $1\leq n \leq d$ so that all states in submanifolds, 
whose dimensions are greater than $(d-n+1)$, are gapped while states are gapless on at least one 
submanifold of dimension $(d-n)$ \cite{benalcazar1,langbehn1,song1,ezawa1,liu1,luo1,kudo1,li1,rchen1}.

\begin{figure}[htbp]
\centering
\includegraphics[width=0.45\textwidth]{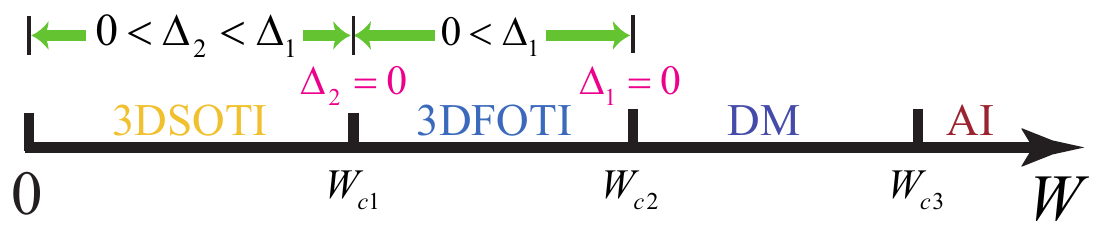}
\caption{A generic route of quantum phase transitions in disordered 3DSOTIs. There are four different phases 
with increasing disorder  $W$: (1) 3DSOTIs characterized by the hinge states localized at the edges; 
(2) 3DFOTIs identified by the surface states; (3) DMs with wave functions spreading over the whole lattice; 
(4) AIs where all states are localized. Here, $\Delta_1$ and $\Delta_2$ are gaps of the bulk and topological surface 
states, respectively. $W_{c1}$, $W_{c2}$, and $W_{c3}$ are the three critical disorders separating the four different phases.
}
\label{fig_1}
\end{figure}

All those newly discovered genuine topological phases should survive in disorders that have profound effects 
on electronic structures as demonstrated in the topological Anderson insulators \cite{sqshen,groth1,suying3}. 
The second-order topological insulators, characterized by in-gap topological states in $(d-2)$-dimensional 
boundary, are the current focus in the field because of its experimental realizations in phononics 
\cite{garcia1,xni1,hfan1}, photonics \cite{hassan1}, and circuitry \cite{simhof1}. So far, most of the works 
are on the constructions of second-order topological insulators in crystal with well-defined crystalline 
symmetries, the fate of such a phase under disorder has been less studied \cite{araki1,su1,agarwala1,szabo1}. 
Thus it should be very interesting to find out how the hinge states in 3DSOTIs are modified by disorders. 
From the knowledge of Anderson localization for topologically trivial states, it is known that the competition 
between the energy randomness and kinetic energy (the bandwidth) determines the metal-to-insulator transition. 
For 3DSOTIs, the gap of bulk states $\Delta_1$ and the gap of topological surface states $\Delta_2<\Delta_1$ 
should be important because a transition from topological surface states to topological states on hinges, 
boundaries of surfaces, can only happen when the surface state gap closes at the transition point. 
We expect that a 3DSOTI undergoes three phase transitions involving four phases as disorder strength 
$W$ increases. The 3DSOTI remains stable up to a critical disorder $W_{c1}$ at which the gap $\Delta_2$ 
of surface states closes ($\Delta_2=0$) while bulk gap remains open ($\Delta_1\neq 0$). The system enters 
into the 3DFOTI from the 3DSOTI. Further increase of disorder to the second critical value of $W_{c2}$ at 
which bulk gap closes, the 3DFOTI is replaced by the conventional topologically trivial diffusive metal (DM). 
The third quantum phase transition is expected to occur at a strong critical disorder $W_{c3}$ at which all 
states are localized and the system becomes an Anderson insulator (AI). The route of these quantum phase 
transitions is presented schematically in Fig.~\ref{fig_1}.
\par

In this work, we use a 3DSOTI model of class A with a reflection-symmetry in the ten Altland-Zirnbauer 
classification \cite{AZclass} to verify the generic route depicted in Fig.~\ref{fig_1}. Using highly-accurate 
numerical calculations, we show that above three quantum phase transitions occur indeed when disorder strength 
$W$ increases. The 3DSOTI is featured by the quantized Hall conductance exactly at $e^2/h$ and constant 
density of states. The 3DFOTI is featured by its dominated occupation probability on surfaces and negligible 
occupation probabilities in the bulk and on hinges. The DMs and AIs are identified from the scaling analysis 
of the participation ratios (PRs), defined as $p_2(E,W)=\langle(\sum_{\bm{i}}|\psi_{\bm{i}}(E)|^4)^{-1}\rangle$ 
with $|\psi_{\bm{i}}(E)|$ being the normalized wave function amplitude at site $\bm{i}$. 
The convincing numerical results are also confirmed by the self-consistent Born approximation (SCBA) calculations.  
\par

This paper is organized as follows. The tight-binding model of 3DSOTI is introduced in Sec.~\ref{sec2}. 
Sec.~\ref{sec3} demonstrates the existence of chiral hinge states in the clean limit. 
Various numerical results are given in Sec.~\ref{sec4} to demonstrate the route of quantum phase transitions in 
Fig.~\ref{fig_1} for a specific set of parameters. In Sec.~\ref{sec5}, a general phase diagram in the plane 
of the Dirac mass $M$ and the disorder strength $W$ is given, followed by the conclusion in Sec.~\ref{sec6}.
\par

\section{Tight-binding Model}
\label{sec2}

Our model is non-interacting electrons on a cubic lattice of lattice constant $a=1$ \cite{langbehn1}
\begin{equation}
\begin{gathered}
H=\sum_{\bm{i}}c^\dagger_{\bm{i}}\left(v_{\bm{i}}\Gamma^0+M\Gamma^2+B\Gamma^{31}\right)c_{\bm{i}}+\\ 
\left(\dfrac{t}{2}\sum_{\langle \bm{ij}\rangle}c^\dagger_{\bm{i}}\Gamma^2 c_{\bm{j}}+\dfrac{it}{2}\sum_{\bm{i}}\left(c^
\dagger_{\bm{i}+\hat{x}}\Gamma^4+c^\dagger_{\bm{i}+\hat{y}}\Gamma^1+c^\dagger_{\bm{i}+\hat{z}}\Gamma^3\right)c_{\bm{i}}+ 
H.c. \right),
\end{gathered}\label{eq1}
\end{equation}
where $c^\dagger_{\bm{i}}\equiv(c^\dagger_{\bm{i}1\uparrow},c^\dagger_{\bm{i}2\uparrow},c^\dagger_{\bm{i}1\downarrow},
c^\dagger_{\bm{i}2\downarrow}$) and $c_{\bm{i}}$ are the electron creation and annihilation operators at site $\bm{i}=(n_x,n_y,n_z)$ 
for orbits 1 and 2, spin up and spin down. $M$ is the Dirac mass that controls the band inversion, and $B$ is a parameter for controlling gap 
opening on surfaces. $\Gamma^0$ and $\Gamma^{\mu=1,2,3,4,5}$ are, respectively, the four-by-four identity matrix and the five non-
unique Dirac matrices satisfying $\{\Gamma^{\mu},\Gamma^{\nu}\}=2\delta_{\mu,\nu}\Gamma^0$ and $\Gamma^{\mu\nu}
=[\Gamma^{\mu},\Gamma^{\nu}]/(2i)$. Here we choose $\Gamma^{(1,2,3,4,5)}=(s_1\otimes\sigma_1,s_2\otimes\sigma_1,s_3\otimes
\sigma_1, s_0\otimes\sigma_3,s_0\otimes\sigma_2)$ with the Pauli matrices $s_\mu$ and $\sigma_\mu$ acting on spin and orbital spaces, 
respectively. $t=1$ is chosen as the energy unit. $v_{\bm{i}}$ is a white noise, distributing uniformly in the range of $[-W/2,W/2]$. 

\section{Clean case}
\label{sec3}

\begin{figure}[htbp]
\centering
\includegraphics[width=0.45\textwidth]{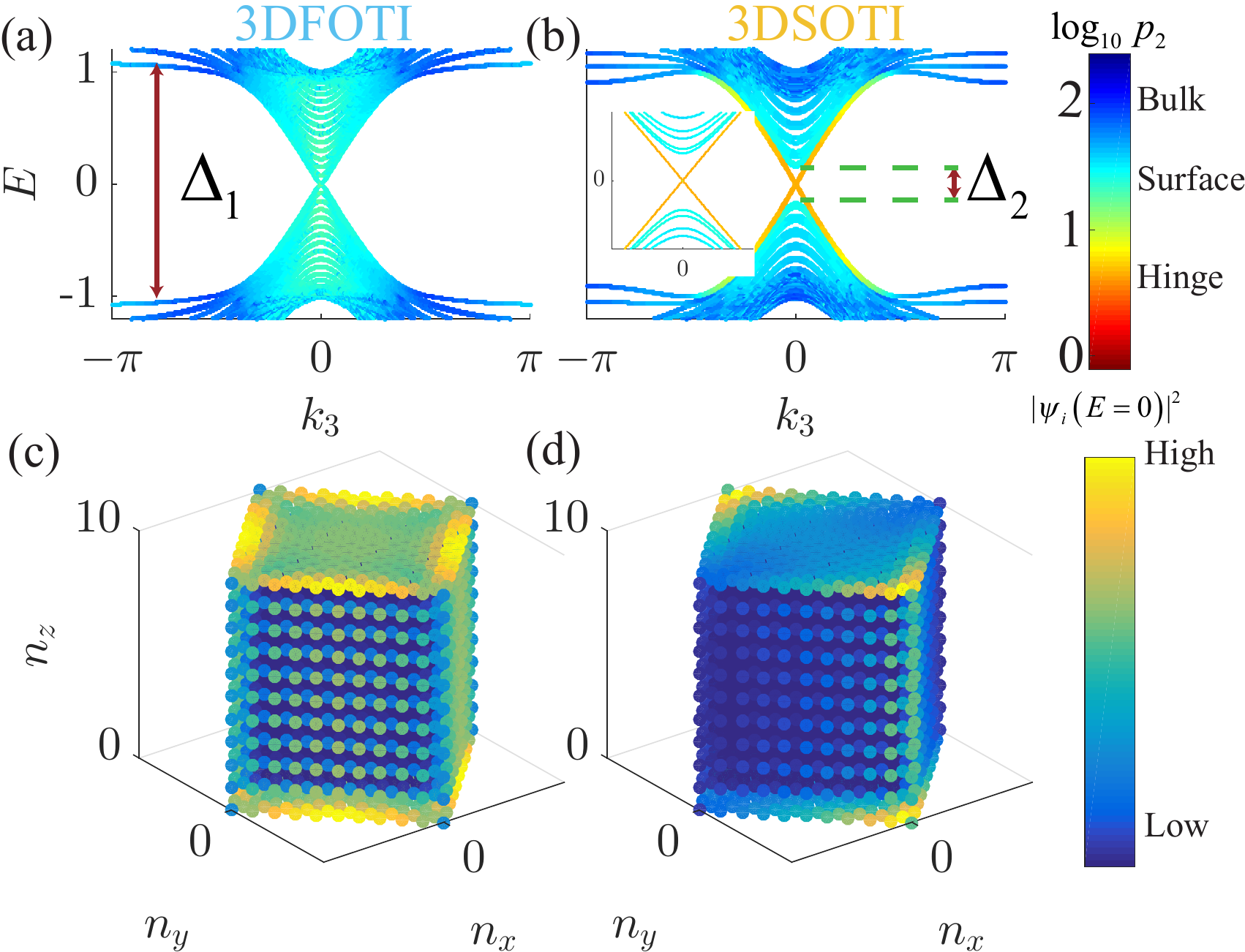}
\caption{(a) $E(k_3)$ for $L=32$, $M=2$, $B=0$ (a) and $B=0.2$ (b). PBC is applied along $z$ direction and OBCs are used on the $(110)$ and $(1\bar{1}0)$ surfaces. Color map $\log_{10}p_2$. The orange/red, cyan/green, and blue color are, respectively, for surface, hinge, and bulk states as denoted by the color bar. Inset in (b) is the enlargement of the zero-energy regime. (c,d) Spatial distribution of wave function (OBCs on all surfaces) of $E=0$ state, $|\psi_{i}|^2=\sum^4_{p=1}|\psi_{i,p}|^2$ for $L=10$ and $B=0$ (c) and $B=0.2$ (d).}
\label{fig_2}
\end{figure}

In the absence of disorder ($W=0$), Hamiltonian~\eqref{eq1} was well studied \cite{langbehn1,trifunovic1} and can be block diagonalized in the momentum space, $H=\sum_{\bm{k}}c^\dagger_{\bm{k}}h(\bm{k})c_{\bm{k}}$ with
\begin{equation}
\begin{gathered}
h(\bm{k})=\sum^4_{\mu=1}d_{\mu}(\bm{k})\Gamma^{\mu}+B\Gamma^{31}.
\end{gathered}\label{eq2}
\end{equation}
Here $d_1(\bm{k})=t\sin k_2$, $d_2(\bm{k})=M-t\sum_{i=1,2,3}\cos k_i$, $d_3(\bm{k})=t\sin k_3$, and $d_4(\bm{k})=t\sin k_1$. For 
$B=0$ and $1<M<3$, Eq.~\eqref{eq2} describes a reflection-symmetric strong 3DFOTI with reflection plane on $x=0$ 
\cite{fu1}. The Hamiltonian does not change under the reflection symmetry of $\Gamma^{54}=s_0\otimes\sigma_1$, i.e., $
\Gamma^{54}h(k_1,k_2,k_3)\Gamma^{54}=h(-k_1,k_2,k_3)$. According to the bulk-boundary correspondence, the non-trivial bulk 
topology guarantees the appearance of the gapless surface states on the self-reflected surfaces, e.g., $H_{\text{surface}}
=v_1k_1\sigma_3+v_2 k_2\sigma_1$ on $z=0$, while the gapped surface states on the non-reflected surfaces
\cite{teo1,chiu1,morimoto1,shiozaki1}. If two non-reflection surfaces encounter at the reflection plane under a sharp angle, the last term $B
\Gamma^{31}$ leads to the band inversion of surface states, as well as the emergence of hinge states at their boundary 
\cite{langbehn1,trifunovic1}. Thus, Hamiltonian~\eqref{eq1} supports a 3DSOTI.
\par

To visualize the above descriptions, we plot the energy spectrum $E(k_3)$ of a rectangle bar sample with open boundary conditions 
(OBCs) on the surfaces perpendicular to $(110)$ and $(1\bar{1}0)$ and periodic boundary condition (PBC) along the $z$ direction 
for $M=2,B=0$ (Fig.~\ref{fig_2}(a)) and $M=2,B=0.2$ (Fig.~\ref{fig_2}(b)), where the colors encode the information of the common 
logarithmic of the participation ratio $p_2$. With the help of $\log_{10}p_2$, one can easily identify the hinge ($\log_{10}p_2<1$), 
surface ($\log_{10}p_2\sim 1$), and bulk states ($\log_{10}p_2\sim 2$). Clearly, for $B=0$, topological surface states exist in 
the bulk gap, i.e., $E\in[-\Delta_1/2,\Delta_1/2]$, while for $B=0.2$ surface states are gapped in $E\in[-\Delta_2/2,\Delta_2/2]$ 
and the hinge states emerge in the gap, exactly the same as reported results \cite{langbehn1,trifunovic1}. 
This can also be seen from the distribution of wave function of $E=0$ shown in Fig.~\ref{fig_2}(c,d).
\par

\section{Disorders induced phase transitions}
\label{sec4}

To see how four different phases illustrated in Fig.~\ref{fig_1} appear under disorders, we set $M=2,B=0.2$ and vary $W$. 
Firstly, we demonstrate that 3DSOTIs in clean limit persist to a finite disorder $W_{c1}$ and then become 3DFOTIs for 
$W>W_{c1}$ by calculating the dimensionless conductances and density of states. Then, we show that a second quantum phase 
transition from 3DFOTIs to DMs at a higher disorder $W_{c2}$ through the analysis of the change of $E=0$-state distribution 
on surfaces (explained below) and density of bulk states, which agree well with the SCBA calculations. 
Finally, we illustrate a third quantum phase transition from DMs to AIs happens at $W_{c3}$ that is larger than $W_{c2}$.

\subsection{3DSOTI-to-3DFOTI}

\begin{figure}[htbp]
\centering
  \includegraphics[width=0.45\textwidth]{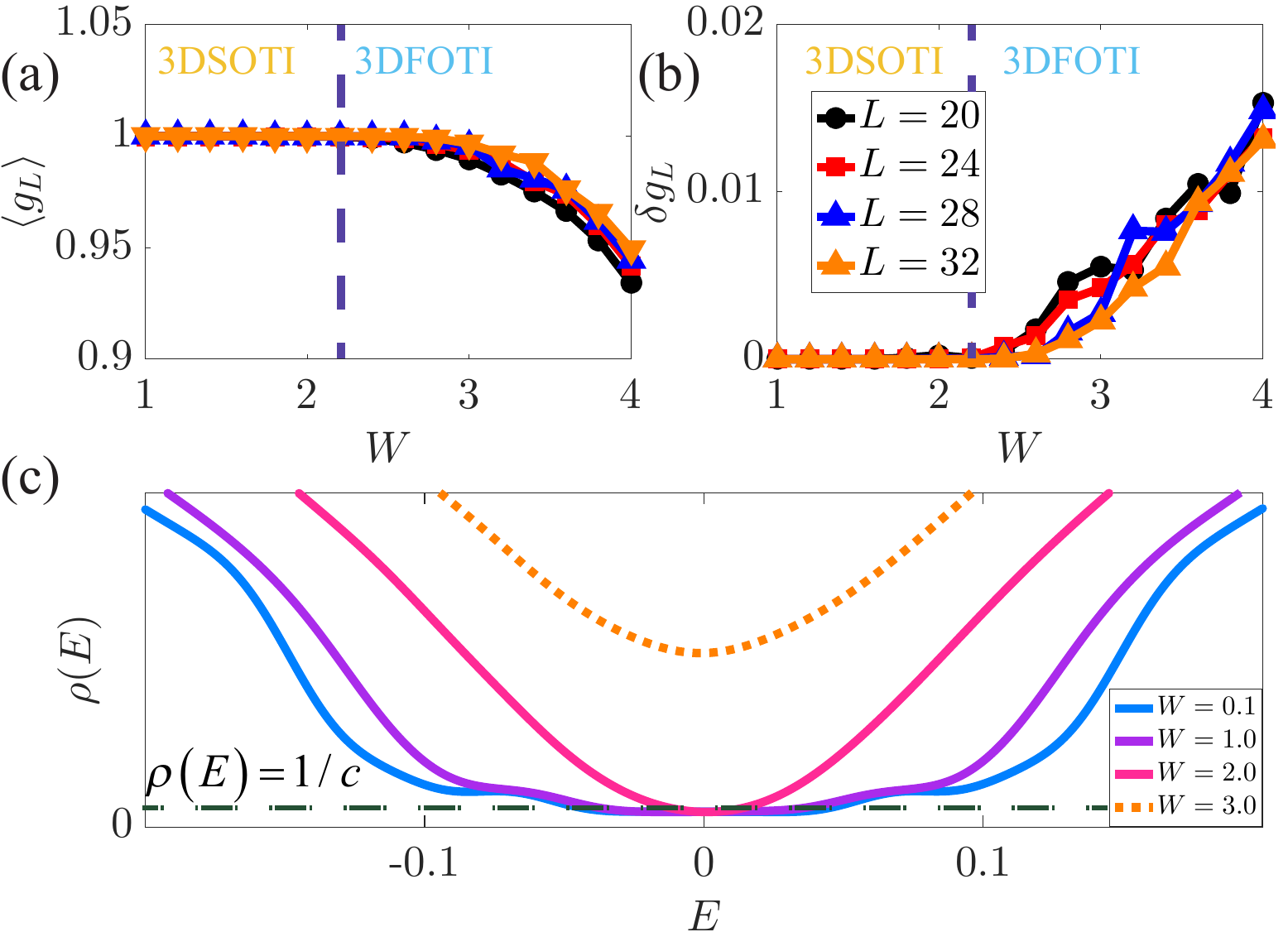}
\caption{(a,b) $\langle g_L\rangle$ (a) and $\delta g_L$ (b) as a function of $W$ for $E=0.02$ and various $L$. 
The dash lines denote $W_{c1}$. (c) $\rho(E)$ for $E\in[-0.15,0.15]$, $L=66$ and various $W$. 
The solid (dotted) lines are for $W\leq W_{c1}$ ($W>W_{c1}$). The dash line guides the eyes for a non-zero 
constant $\rho(E)=1/c$.}
\label{fig_3}
\end{figure}

We consider the Hamiltonian \eqref{eq1} on a bar of size $(L/\sqrt{2}\times L/\sqrt{2}\times L)$. 
The OBCs on the surfaces perpendicular to $(110)$, $(1\bar{1}0)$, and $(001)$ are assumed if not specified otherwise. 
Note that hinge states in 3DSOTIs are chiral, the conductance through them must be quantized with zero fluctuations 
\cite{langbehn1} if there is no other conduction channels exist at the Fermi level. This feature can be used to distinguish 
a 3DSOTI hinge state from others. Thus, we compute the two-terminal dimensionless conductances $g_L$ of a disordered bar 
connected to two semi-infinite leads along $z$ direction. The dimensionless conductance is given by $g_L=\text{Tr}[TT^\dagger]$ 
with $T$ being the transmission matrix \cite{macKinnon1}. The Fermi energy is fixed at $E=0.02$ to focus on the hinge states. 
Figures~\ref{fig_3}(a) and (b) show, respectively, the sample-averaged dimensionless conductance $\langle g_L\rangle$ and the 
conductance fluctuation $\delta g_L=(\langle g^2_L\rangle-\langle g_L\rangle^2)^{1/2}$ as a function of disorder $W$ for 
various sizes from $L=20$ to $L=32$. Clearly, all $\langle g_L\rangle$ are exactly quantized at 1 for $W<W_{c1}\simeq 2.2$ 
with zero conductance fluctuation, a typical feature of hinge states. Beyond $W_{c1}$, $\langle g_L\rangle$ is not quantized 
and $\delta g_L\neq 0$. $\langle g_L\rangle$ increases with $L$, an indication of states of $E=0.02$ being extended. 
As demonstrated later, these states are surface states, and the system is a 3DFOTI.   
\par

The dispersion relation of the hinge states in clean 3DSOTIs is linear in $k_3$, i.e., $E_{\text{hinge}}=\pm ck_3$, 
see Fig.~\ref{fig_2}(b). Thus its contribution to the density of states (DOS) is a constant, i.e., $\rho(E)=1/c$. 
Interestingly, the average DOS of disordered 3DSOTI, defined as $\rho(E)=\langle 1/(4L^3)\sum_{q}\sum^4_{p=1}
\delta(E-E_{p,q})\rangle$ and obtained from the kernel polynomial method \cite{kpm} is a disorder-independent constant. 
Average $\rho(E)$ of the disordered bar of $L=66$ for various $W$ are plotted in Fig.~\ref{fig_3}(c). 
Apparently, the width of constant DOS becomes smaller as $W$ increases. This is expected since disorders tend to 
reduce the gap $\Delta_2$. For large enough disorders $W>W_{c1}$, the constant plateau of $\rho(E)$ disappears when 
the gap $\Delta_2$ of surface state vanishes and the system becomes a 3DFOTI. 
In summary, constant DOS is another fingerprint of the disordered 3DSOTI.
\par

\subsection{3DFOTI-to-DM}

After establishing the existence of 3DSOTIs for $W<W_{c1}$, we would like to show now that the system is a 
3DFOTI for $W\in[W_{c1},W_{c2}]$, where the zero-energy states are the surface states rather than the hinge states,
and becomes a DM for $W>W_{c2}$. Note that the wave function of $E=0$ of both 3DFOTIs and 3DSOTIs are highly 
localized at system boundaries, either on the surfaces in a 3DFOTI or on hinges in a 3DSOTI, in contrast to be 
extended over the whole systems in a DM. Therefore, we use the following quantity to distinguish states in  
3DSOTIs and 3DFOTIs from those in DMs:
\begin{equation}
\begin{gathered}
\zeta_{W,L}=\sum_{i\in\text{Surface}}\sum^4_{p=1}
|\psi_{i,p}(E=0)|^2
\end{gathered}\label{eq4}
\end{equation}
with the first summation over all sites on the surfaces. $\zeta_{W,L}$ describe the wave function distribution of 
the zero energy states on surfaces. As expected, $\zeta_{W,L=\infty}$ is a finite non-zero constant for 3DSOTIs 
and approaches zero for a DM since the ratio of number of surface sites to that of bulk sites goes to zero. For a 
fixed $W$, $\zeta_{L,W}$ should increase with $L$ for 3DFOTIs and 3DSOTIs and decreases with $L$ for DM. 
They should intersect at critical disorder $W_{c2}$.  Numerically, we use the retarded Lanczos method to find the 
eigenfunction of the level nearest to $E=0$ of the disordered bar and calculate $\zeta_{W,L}$. 
In our scenario, we first use the KWANT package \cite{kwant} to construct a Hamiltonian matrix $H$ out of 
tight-binding model Eq.~\eqref{eq1}. We then solve the eigenequation $H\psi=E\psi$ using the SCIPY library 
\cite{scipy} to obtain the required eigenenergies and eigenfunctions.
\par

\begin{figure}[htbp]
\centering
\includegraphics[width=0.45\textwidth]{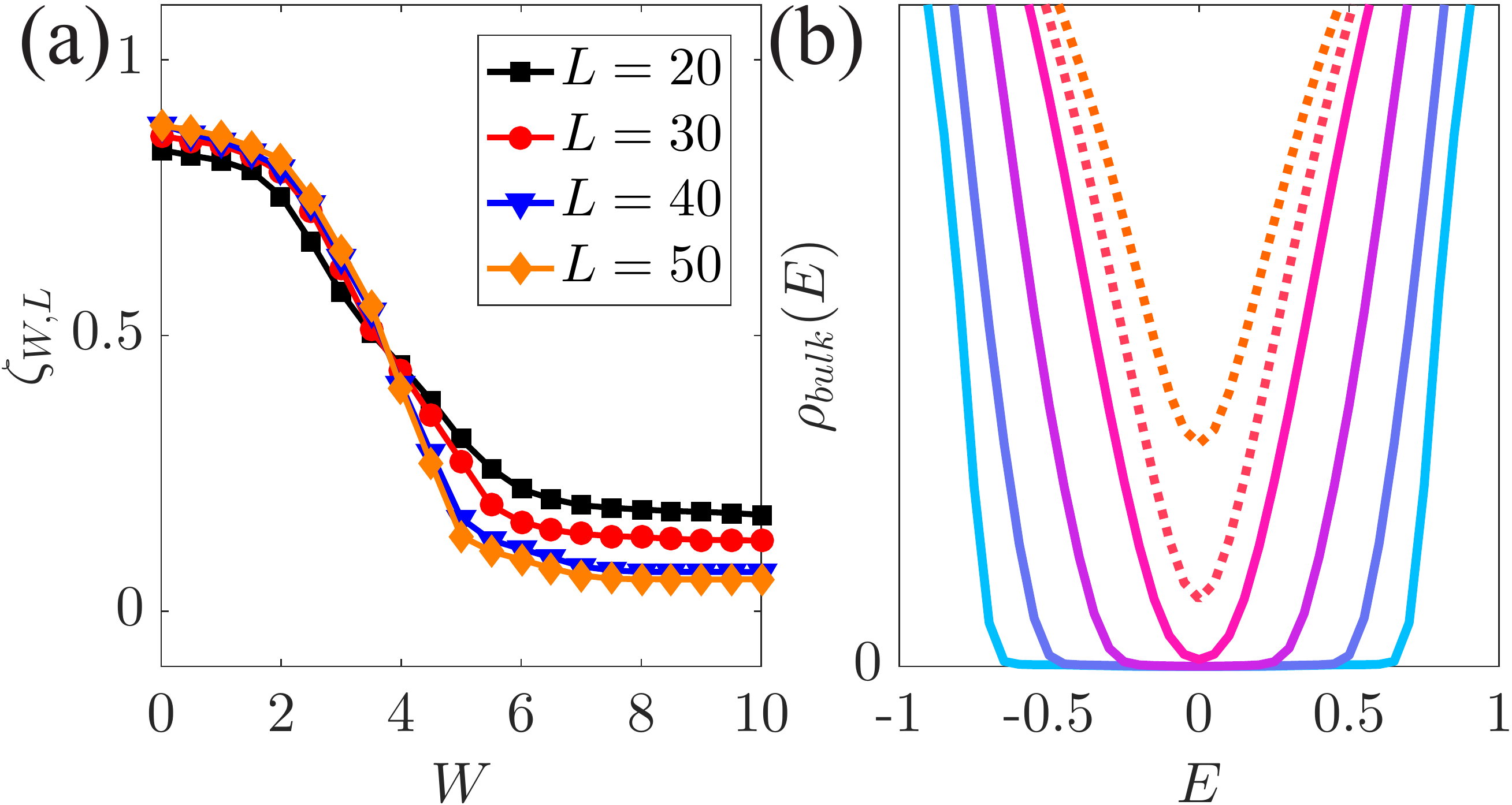}
\caption{(a) $\zeta_{W,L}$ v.s. $W$ for various $L$. (b) $\rho_{\text{bulk}}(E)$ for various $W$ (the solid and 
the dash lines are for $W=1,2,3,4<W_{c2}$ and $W=4.5,5>W_{c2}$, respectively).}
\label{fig_4}
\end{figure}

To substantiate the above assertion, we evaluate the sample-averaged surface density $\zeta_{W,L}$ for various $L$'s 
ranging from $20$ to 50 and various $W$. The results are plotted in Fig.~\ref{fig_4}(a). A phase transition, between 
the boundary states ($d\zeta_{W,L}/dL>0$) and the bulk states ($d\zeta_{W,L}/dL<0$) at $W=W_{c2}\simeq 4$ at which 
all $\zeta_{W,L}$ curves cross, can be clearly seen. Since $W_{c2}>W_{c1}$, the 3DFOTIs, where wave function of $E=0$ 
is localized on the surface rather than the edge of the bar as illustrated in Appendix~\ref{app1}, exist between the 
3DSOTIs at weak disorders $W<W_{c1}$ and the DMs beyond $W_{c2}$ ($W>W_{c2}$). Figure~\ref{fig_4}(a) is thus a 
verification of the existence of 3DFOTI-to-DM quantum phase transitions such that $E=0$ states for $W_{c1}<W<W_{c2}$ 
and $W>W_{c2}$ belong, respectively, to the 3DFOTIs and the DMs in the thermodynamic limit of $L\to\infty$.
\par

The 3DFOTI-DM transition happens when the bulk gap closes. To substantiate it, we calculate $\rho_{\text{bulk}}(E)$ 
for the disordered bar of $L=66$ with PBCs along all direction so that all boundary states (surfaces or hinges) are 
eliminated, and only bulk states can contribute to DOS. The DOS for various disorders are displayed in Fig.~\ref{fig_4}(b). 
As expected, $\rho_{\text{bulk}}(0)=0$ below the critical disorder of $W_{c2}\simeq 4.0$ and $\rho_{\text{bulk}}(0)\neq 0$ 
beyond $W_{c2}$, in contrast to non-zero $\rho(E)$ around $E=0$ in Fig. \ref{fig_3}(c) for $W<W_{c2}$. 
Figure~\ref{fig_4}(a) demonstrates from a different angle that non-zero $\rho(E)$ around $E=0$ in Fig. \ref{fig_3}(c) is from either 
hinge states of 3DSOTI or surface states of 3DFOTI. The estimate of the critical disorder strength is consistent with 
that by the finite-size analysis of $\zeta_{W,L}$.  
\par

We carry out the SCBA calculations to further understand the disorder effects \cite{groth1,czchen1,sliu1,suying3}. 
The self energy is $\Sigma(E)=(W^2/12N)\sum_{\bm{k}}((E+i0)\Gamma^0-h(\bm{k})-\Sigma(E))^{-1}$, where $N$ is the 
total number of sites. For simplicity, the $B\Gamma^{31}$ term ($B\ll M$ in this work) is neglected and $\Sigma$ can 
be expressed as $\Sigma=\sum^4_{\mu=0}\Sigma_{\mu}\Gamma^{\mu}$. For $E=0$, $\Sigma_0=-i(1/\tau)$ is a pure imaginary 
number, with $\tau$ being the lifetime of the zero-energy bulk states, i.e., $\rho_{\text{bulk}}(E=0)\propto(1/\tau)$. 
After some algebra (see Appendix~\ref{app2}), we obtain
\begin{equation}
\begin{gathered}
\dfrac{1}{\tau}=\dfrac{1}{\tau}\dfrac{W^2}{12N}
\sum_{\bm{k}}
\dfrac{1}{\sum^4_{\mu=1}(d_\mu(\bm{k})+\Sigma_\mu(0))^2+(1/\tau)^2}
\end{gathered}\label{eq6}
\end{equation}
with $\Sigma_{1,3,4}=0$. Here, the summation is taken over the first Brillouin zone (BZ). 
The solutions of Eq.~\eqref{eq6} can be either $1/\tau=0$ for $W\leq W_{c2}$ or $1/\tau\neq 0$ for $W>W_{c2}$. 
The former corresponds to either 3DFOTIs or 3DSOTIs where $\Delta_1\neq 0$ and $\rho_{\text{bulk}}(0)=0$, while 
the later is for DMs with $\Delta_1=0$ and $\rho_{\text{bulk}}(0)\neq 0$. The critical disorder $W_{c2}$ is 
given by the gap equation \cite{sliu1}
\begin{equation}
\begin{gathered}
1=\dfrac{W^2_{c2}}{12N}
\sum_{\bm{k}}
\dfrac{1}{\sum^4_{\mu=1}(d_\mu(\bm{k})+\Sigma_\mu(0))^2}.
\end{gathered}\label{eq7}
\end{equation}
Numerically, we obtain $W_{c2}\simeq 3.7$ for $M=2$ and $B=0.2$, consistent with the estimates from $\zeta_{W,L}$ 
and $\rho_{\text{bulk}}(E)$. According to the SCBA, $M$ is renormalized by the disorder as $\tilde{M}=M+\Delta$ with 
\begin{equation}
\begin{gathered}
\Delta=-\dfrac{W^2}{12N}\sum_{\bm{k}}\dfrac{d_2(\bm{k})+\Delta}
{\sum^4_{\mu=1}(d_\mu(\bm{k})+\Sigma_\mu(0))^2}
\end{gathered}\label{eq9}
\end{equation}
such that the phase boundary between 3DSOTIs, 3DFOTIs and topological trivial phase are shifted by disorders.
\par

\subsection{DM-to-AI}

The extended states in DMs are eventually localized by strong disorders. To investigate the nature of this 
Anderson localization transition and its associated universality class, we compute the PR $p_2(E=0,W)$, which 
measures how many lattice sites are occupied by the wave function of $E=0$ \cite{janssen1,xrwang1,cwang1}. 
Near the critical disorder  $W_{c3}$ of the Anderson localization transitions, $p_2$ satisfies the one-parameter 
scaling function \cite{pixley1,cwang2}
\begin{equation}
\begin{gathered}
p_2(W)=L^D [f(L/\xi)+CL^{-y}],
\end{gathered}\label{eq8}
\end{equation}
where $f(x)$ is the unknown scaling function, $C$ and $y>0$ are a constant and the exponent of the irrelevant 
variable, respectively. The correlation length $\xi$ diverges at $W_{c3}$ as $\xi\propto|W-W_{c3}|^{-\nu}$ with 
critical exponent $\nu$. $D$ is the fractal dimension of critical wave functions which occupy a subspace of 
dimensionality smaller than the embedded space dimension $d=3$. By defining $Y_L(W)=p_2L^{-D}-CL^{-y}$, 
we use the following criteria to identify an Anderson localization transition \cite{cwang2}: 
(1) $dY_L(W)/dL>0$ ($dY_L(W)/dL<0$) for DMs (AIs). (2) Near $W_{c3}$, $Y_L(W)$ of different $L$ collapse 
into a single curve of $f(x)$.
\par

\begin{figure}[htbp]
\centering
  \includegraphics[width=0.45\textwidth]{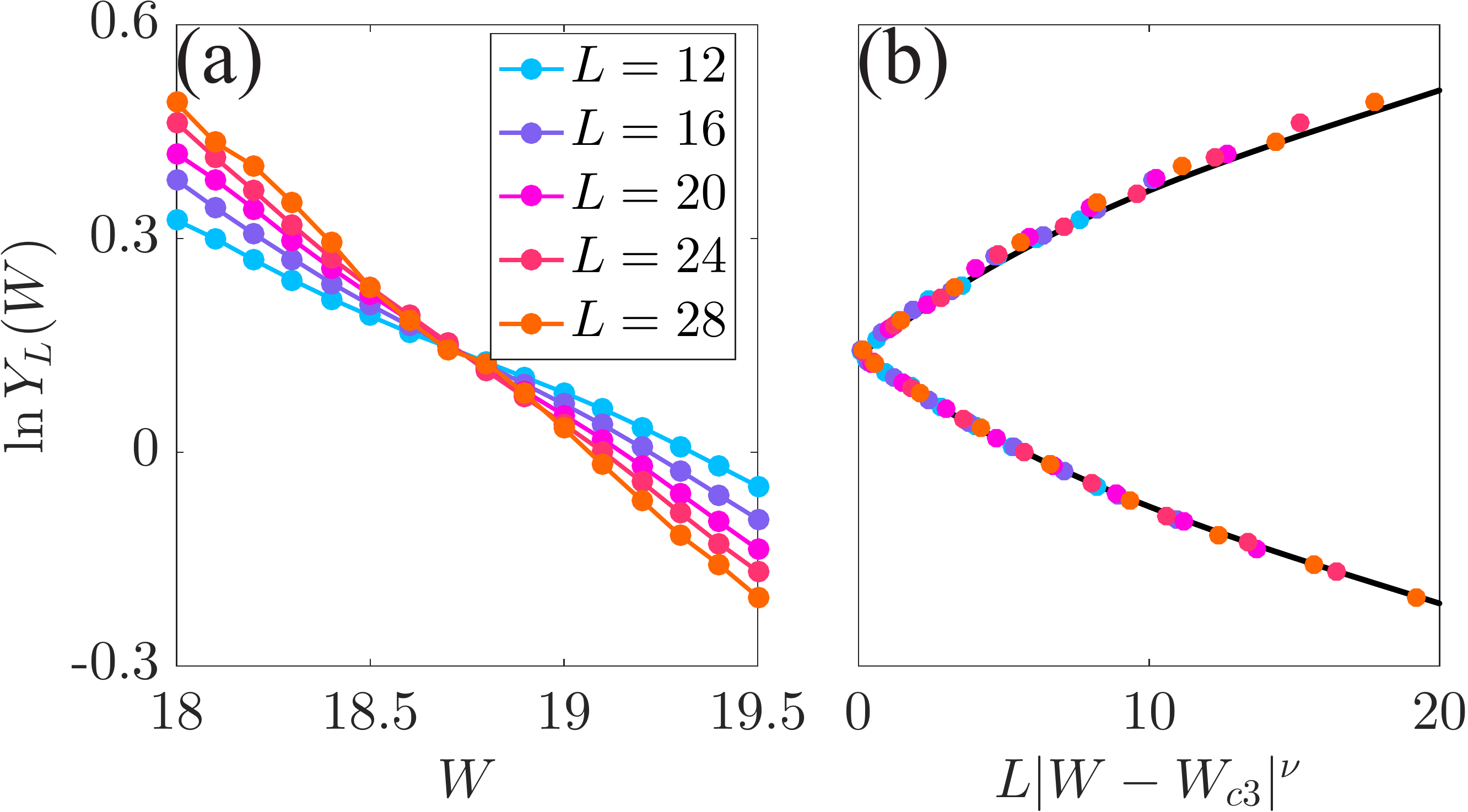}
\caption{(a) $\ln Y_{L}(W)$ for $E=0$ and various $L$. (b) Scaling function $\ln Y_L(W)=\ln f(x=L|W-W_{c3}|^{\nu})$ 
for the Anderson localization transition.}
\label{fig_5}
\end{figure}

Near $W_{c3}$, the calculated $\ln Y_{L}(W)$ and $\ln f(x)$ are displayed in Figs.~\ref{fig_5}(a) and (b), respectively. 
Data in Fig.~\ref{fig_5}(a) give $W_{c3}=18.73\pm 0.03$, and $dY_L(W)/dL$ is always positive (negative) for $W<W_{c3}$ 
($W>W_{c3}$), indicating the system is a DM (AI). Following the well-established procedure \cite{cwang2}, we find 
$D=1.7\pm 0.2$, and $\nu=1.5\pm 0.2$ (see Appendix~\ref{app3} for more details). The obtained $\nu$ and $D$, 
characterizing the universality class of transitions, are consistent with previous estimations for 3D Gaussian unitary 
ensemble \cite{kawarabayashi1}.   
\par

\section{Phase diagrams}
\label{sec5}

\begin{figure}[htbp]
\centering
\includegraphics[width=0.45\textwidth]{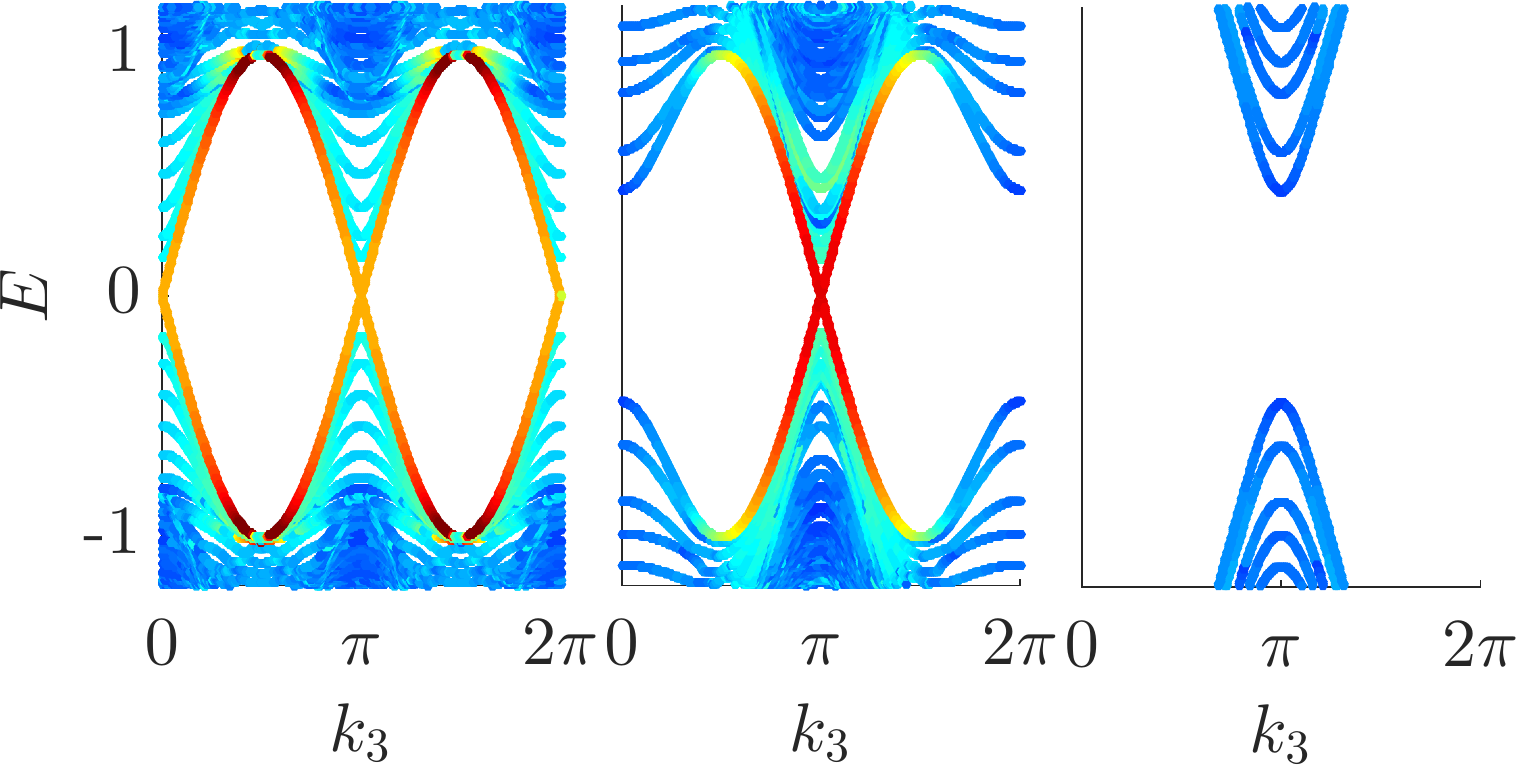}
\caption{Energy spectrum $E(k_3)$ for $L=32$, $B=0.2$ and three typical $M$. Left: $M=0$ (3DSOTI); middle: 
$M=1.5$ (3DSOTI); right: $M=3.5$ (normal insulator). 
Colors encode $\log_{10}p_2$ and color bar is the same as those in Figs.~\ref{fig_2}(a,b).} 
\label{fig_6}
\end{figure}

\begin{figure}[htbp]
\centering
  \includegraphics[width=0.42\textwidth]{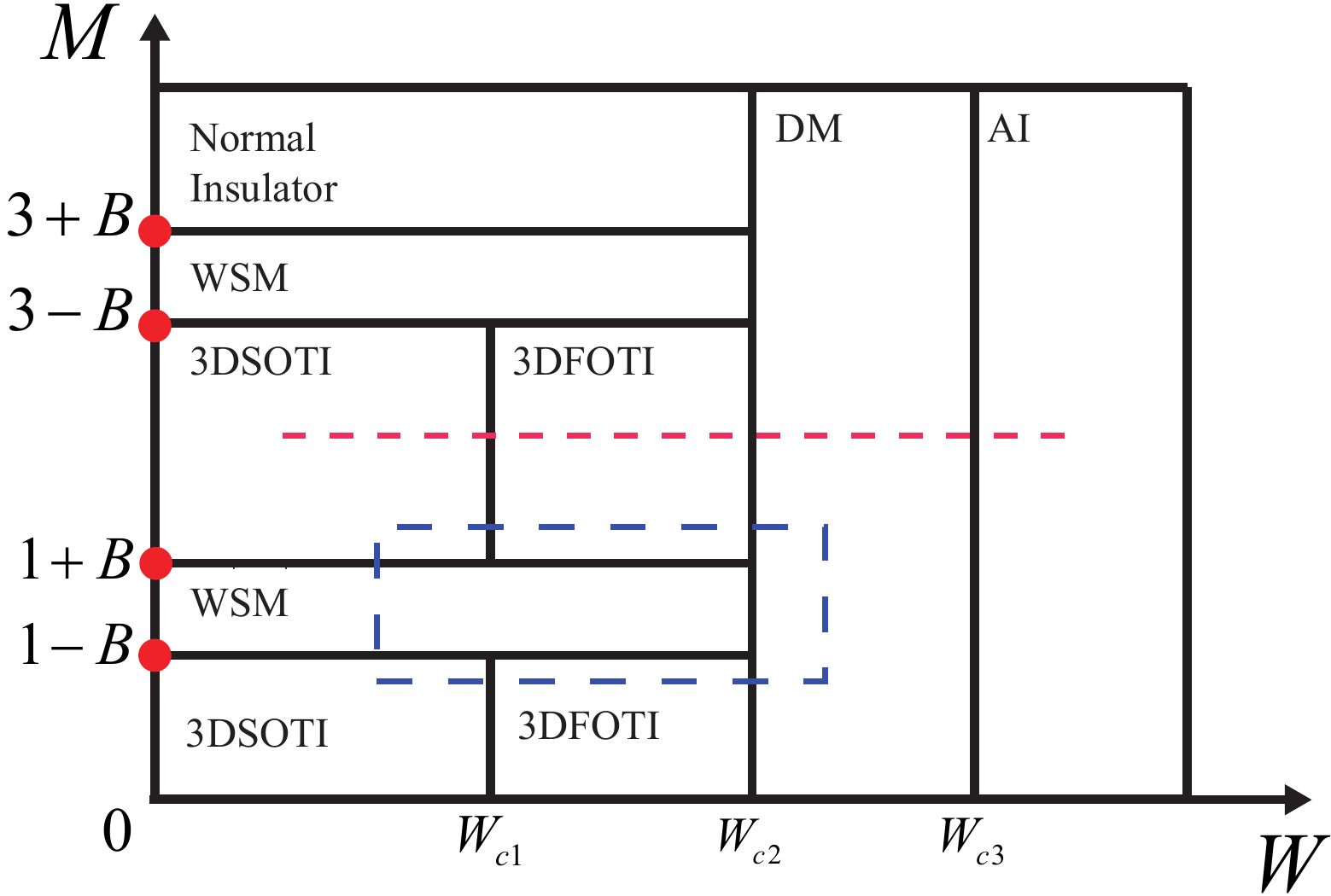}
\caption{Schematic of the general phase diagrams in $W-M$ plane. $M$ is the Dirac mass that drives quantum phase 
transitions between 3DFOTIs, 3DSOTIs, WSMs, and normal insulators. $W$ measures the disorder strength. $B$ is 
the tuning parameter for the term that causes the band inversions in surface states of 3DFOTIs. We only consider 
$B=0.2<1$ here. The phase boundaries are the sketches only.}
\label{fig_7}
\end{figure}

We would like to depict a more inclusive general phase diagram of the model by varying both $M$ and $W$ for a fixed 
$B=0.2$ in this section, in contrast of varying $W$ only for fixed $M$ and $B$. Let us discuss the clean case first. 
For $M>0$, there are four critical points at $M=1\pm B$ and $3\pm B$ that separate three different phases from each 
other (see the red dots in Fig.~\ref{fig_7}). For $M\in[1-B,1+B]$ ($M\in[3-B,3+B]$), Eq.~\eqref{eq2} is a gapless 
Weyl semi-metal (WSM) with three (one) pairs of Weyl nodes in the first BZ, while the band gap is non-zero for 
$0<M<1-B$, $1+B<M<3-B$, and $M>3+B$ (see Appendix~\ref{app4} for more details). As $M$ increases, we find that the 
system is a 3DSOTI for $0<M<1-B$; enters a WSM at $M=1-B$; re-enters into the 3DSOTI at $M=1+B$; becomes a WSM again 
for $3-B<M<3+B$; and is a normal gapped insulator for $M>3+B$ where there is no band inversion for surface states. 
Remarkably, there are two pairs of one-dimensional helical edge channels (hinge states) for $M<1-B$, indicating the occurrence 
of band inversions for two Dirac cones. Consequently, the two-terminal conductance should be quantized at $2e^2/h$ for 
$M<1-B$. On the other hand, there is only one surface Dirac cone for $1+B<M<3-B$, the conductance of the 3DSOTI is 
thus quantized to $e^2/h$. The one or two Dirac cones show clearly in the spectrum $E(k_3)$ of a Hall bar with OBCs 
on the surfaces perpendicular to $(110)$ and $(1\bar{1}0)$ and PBCs along the $z-$direction. 
Figs.~\ref{fig_6}(a,b,c) plot $E(k_3)$ respectively for $M=0,1.5,3.5$. 

Naturally, we expect that the 3DSOTIs, due to the band inversion of surface states with both one or two Dirac 
cones shown in Figs.~\ref{fig_6}(a,b), are robust against weak disorders. With the increase of $W$, a transition 
from a 3DSOTI to a 3DFOTI occurs at a critical disorder $W_{c1}$ where the gap of surface states $\Delta_2$ closes 
but the gap of bulk states $\Delta_1$ remains open. At some higher critical disorders $W_{c2}$, $\Delta_1=0$ such 
that a transition from 3DFOTIs to DMs occurs. Finally, an Anderson localization transition from DMs to AIs occurs at 
large enough disorders $W_{c3}$. All these features can be seen from a schematic phase diagram in Fig.~\ref{fig_7}. 
In addition to the four phases shown in Fig.~\ref{fig_1}, there are two more phases: the WSMs characterized by 
paired Weyl nodes in clean limit and the normal gapped insulators. Noticeably, the phase boundary of the disordered 
WSMs is still an issue under debate, e.g., whether the WSMs can exist in finite disorders \cite{pixley1} and whether 
there is a direct WSM-to-DM transition without the intermediated Chern insulator phase \cite{czchen1,sliu1,cwang2} 
or two quantum phase transitions of WSM-to-CI-to-DM with increasing disorders \cite{ysu1}. 
However, this challenging problem is not the focus of this work.
\par

\begin{figure}[htbp]
\centering
  \includegraphics[width=0.45\textwidth]{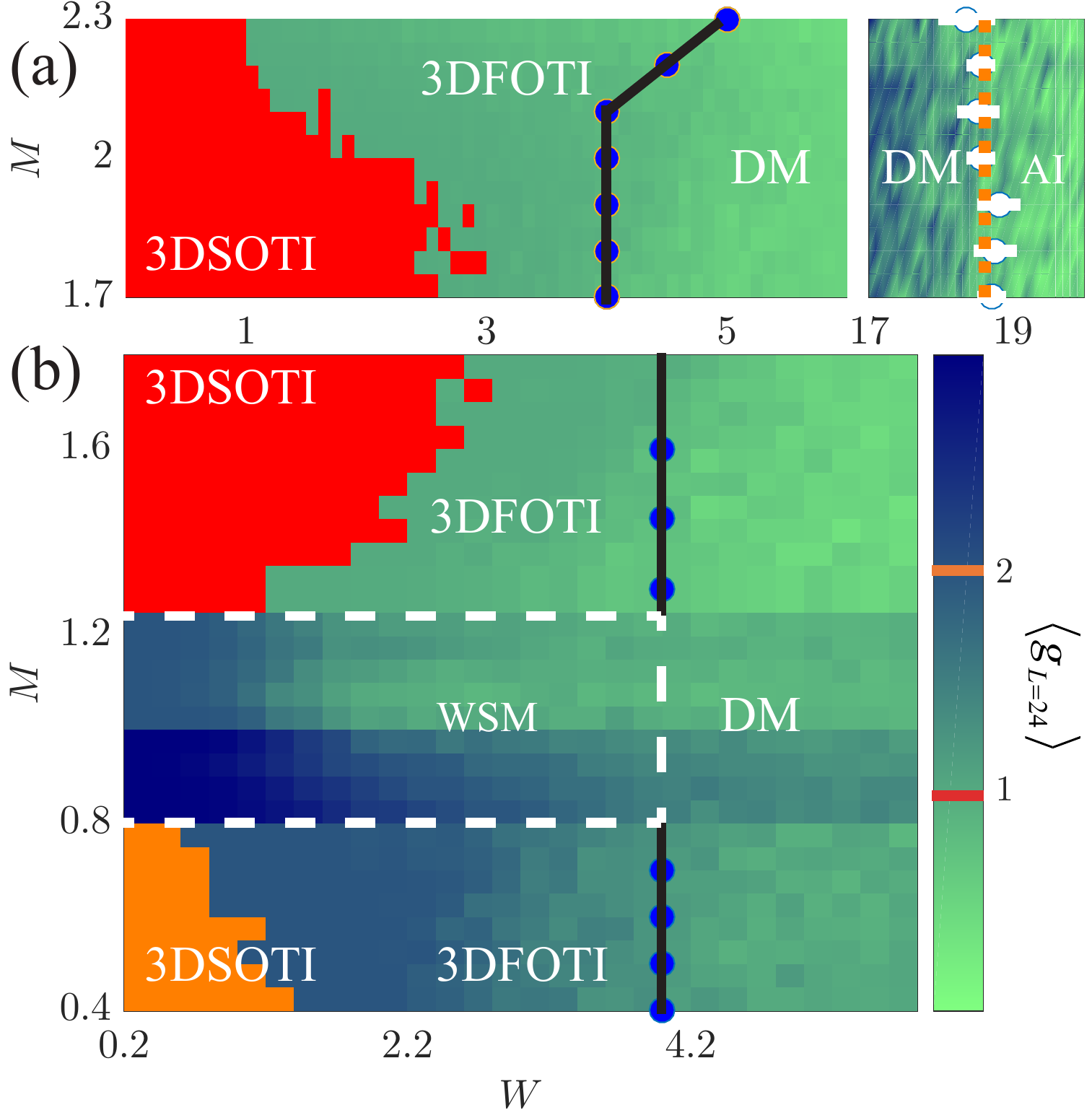}
\caption{Phase diagram in the $W-M$ plane displaying the occurrences of 3DSOTIs, 3DFOTIs, DMs, and AIs for different 
parameter regimes of Fig.~\ref{fig_7}: (a) along the red line; (b) within the blue rectangle. Colors encode 
$\langle g_{L=24}(W,M)\rangle$. 3DSOTIs are characterized by the quantized conductances. The boundaries between 
3DFOTIs and DMs (black solid lines) are determined by finite-size scaling analysis of $\zeta_{W,L}$. The boundary 
between DMs and AIs (orange dash line) is given by scaling analysis of $p_2(W,L)$. White dash lines show the boundary 
of disordered WSMs schematically, which is still controversial.
}
\label{fig_8}
\end{figure}

We have partly confirmed the general phase diagram through exhaustive numerical calculations of different $M$ and $W$ and a fixed 
$B=0.2$. We first consider the phase transitions along the red line in Fig.~\ref{fig_7}. A phase diagram in the $W-M$ plane 
displaying the existences of the 3DSOTI, 3DFOTI, DM, and AI phases is shown in Fig.~\ref{fig_8}(a). 
$W_{c1}$ and $W_{c2}$ respectively and monotonically decreases and increases with $M$ while $W_{c3}$ does not depend on $M$. 
Although the phase boundaries, i.e., $W_{c1}$, $W_{c2}$, and $W_{c3}$, depend on the details of a model, the physics in 
Fig.~\ref{fig_8}(a) and Fig.~\ref{fig_1} is general. 
\par

We provide additional supports to the general phase diagram by probing the blue dash rectangle regime in Fig.~\ref{fig_7}. 
The results are shown in Fig.~\ref{fig_8}(b). As expect, the dimensionless conductance $\langle g_{L=24}\rangle$ are exactly 
quantized at 2 and 1 for the 3DSOTIs with $0<M<1-B$ and $1+B<M<3-B$, respectively. Indeed, it is so as shown by the orange 
and red colors in Fig.~\ref{fig_8}(b). While transitions from 3DSOTIs to 3DFOTIs occurs at critical disorders $W_{c1}$ above 
which the conductances lose the quantization. Further increasing $W$ to $W_{c2}$, the system undergoes a transition from 
3DFOTIs to DMs. The critical disorder $W_{c2}$ is also determined by the scaling analysis of $\zeta_{W,L}$ (probability for 
the electron on surfaces), as we did in Fig.~\ref{fig_3}(a).  
\par

\section{Conclusion}
\label{sec6}

In conclusion, a generic route of disorder-induced phase transitions for a crystal 3DSOTI is revealed. 
As random potential strength increase, the 3DSOTI transforms to a 3DFOTI at a lower weak critical disorder 
of $W_{c1}$, followed by a second transition at an intermediate higher critical disorder of $W_{c2}$ to a DM. 
The system eventually becomes an AI after a metal-to-insulator transition at an even stronger critical disorder of $W_{c3}$. 
The 3DSOTI is featured by quantized conductance at $e^2/h$ and zero conductance fluctuation, as well as the 
constant density of states, while 3DFOTIs are identified by their dominate occupation on surfaces, negligible occupation 
in the bulk and on hinges. The DM and AI are confirmed by the scaling analysis of participation ratios, and the 
corresponding Anderson localization transition belongs to the conventional 3D Gaussian unitary class. 
We believe that such general route should be held for two-dimensional second-order topological insulators too, 
but whether there is an intermediate DM phase depends on system symmetries \cite{cwang1}.
\par

\section{Acknowledgments}

This work is supported by the National Natural Science Foundation of China (Grants No.~11774296, 11704061 and  11974296) 
and Hong Kong RGC (Grants No.~16301518 and 16301619). CW acknowledges the kindly help from Jie Lu and Xuchong Hu.
\par

\appendix

\section{Additional evidence for transitions from 3DFOTI to 3DSOTI}
\label{app1}

\begin{figure}[htbp]
\centering
  \includegraphics[width=0.35\textwidth]{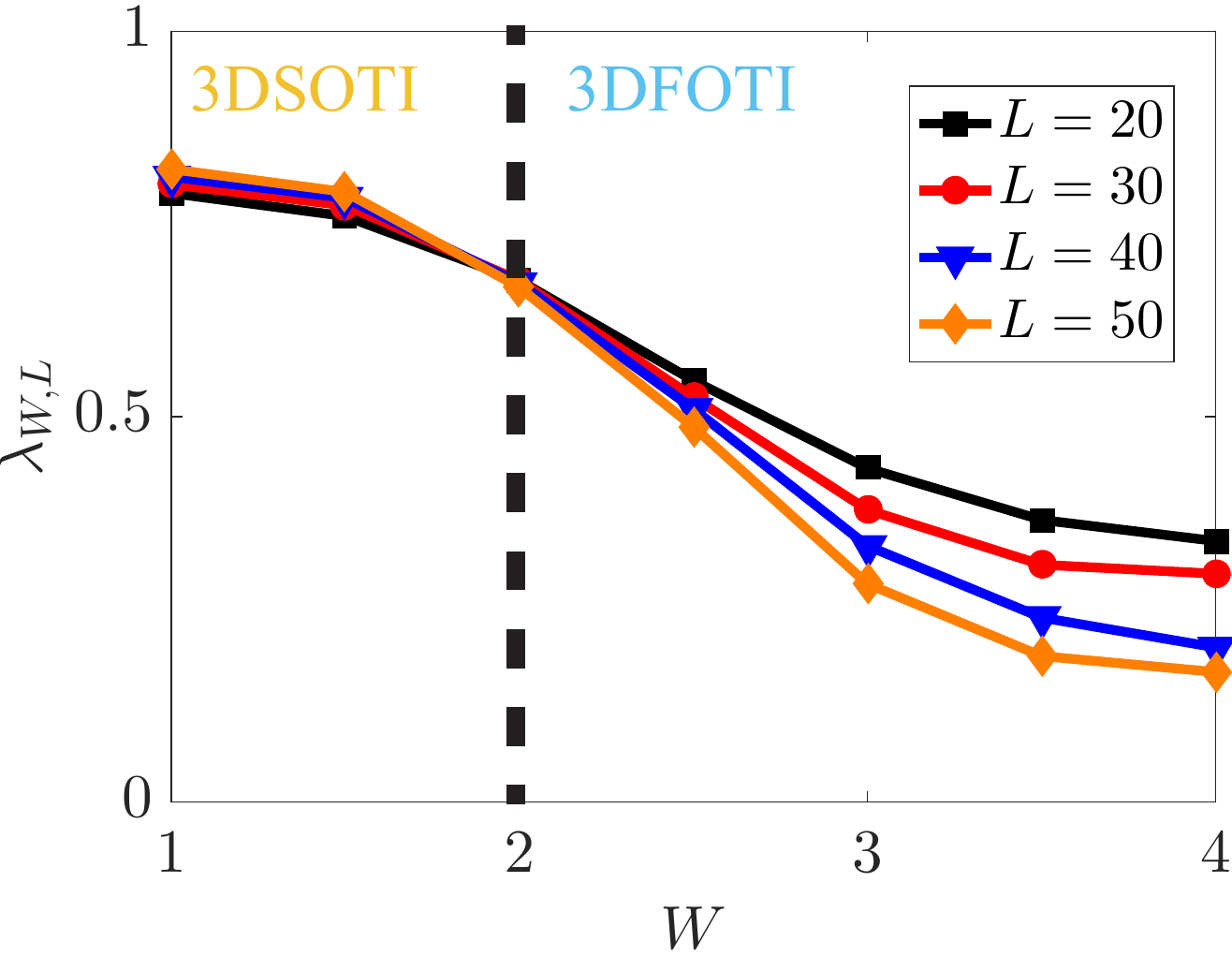}
\caption{$\lambda_{W,L}$ for $M=2,B=0.2$ and various $W$ and $L$. 
The dash line locates the $W_{c1}$.}
\label{fig_a1}
\end{figure}

In this section, we show that the occupation probabilities of an electron in the $E=0$-state lie on the hinges 
(surfaces) if $0<W<W_{c1}$ ($W_{c1}<W<W_{c2}$). We define edge occupation probability $\lambda_{W,L}$ 
\begin{equation}
\begin{gathered}
\lambda_{W,L}=\sum_{i\in\text{edge}}\sum^4_{p=1}|\psi_{i,p}(E=0)|^2,
\end{gathered}\label{add_1_2}
\end{equation}
where $\sum_{i\in\text{edge}}$ is over the two edges of the Hall bars lying on the reflection plane $x=0$. 
The edge occupation probability $\lambda_{W,L}$ is order of 1 for a 3DSOTI and negligible small for a 3DFOTI 
in the limit of $L \rightarrow \infty$. The calculated $\lambda_{W,L}$ for $W=1,1.5,\cdots 4$ and $L=20,30,40,50$ 
are plotted in Fig.~\ref{fig_a1}. Indeed $\lambda_{W,L}$ approaches a finite non-zero number for $W<W_{c1}$ and 
decrease with both $W$ and $L$ for $W>W_{c1}\simeq 2.2$, an obvious feature of transition from a 3DSOTI to a 3DFOTI. 
It is more convincing if one recall the featureless behaviour of surface occupation probability $\zeta_{W,L}$ 
around $W_{c1}\simeq 2.2$ shown in Fig.~\ref{fig_4}(a). This extra property of hinge occupation probability is 
another support of 3DSOTI-to-3DFOTI transition at $W_{c1}$.

\section{Self-consistent Born approximation}
\label{app2}

Within the framework of the self-consistent Born approximation \cite{groth1}, the self-energy $\Sigma$ reads
\begin{equation}
\begin{gathered}
\Sigma(E)=\dfrac{W^2}{12N}\sum_{\bm{q}}\bar{G}(\bm{q},E)
\end{gathered}\label{eq2_9}
\end{equation}
with the summation $\sum_{\bm{q}}$ taking over the first BZ. Now our task is to solve
\begin{equation}
\begin{gathered}
\bar{G}^{-1}(\bm{k},E)=G^{-1}_0(\bm{k},E)-\dfrac{W^2}{12N}\sum_{\bm{q}}\bar{G}(\bm{q},E)
\end{gathered}\label{eq2_10}
\end{equation}
with the free Green function $G_0(\bm{k},E)=((E+i0)\Gamma_0-h(\bm{k}))^{-1}$. In order to avoid a complicated calculation, we omit the $B\Gamma^{31}$ term (since $B\ll 1$) such that the free Green function is the combination of $\Gamma^0$ and $\Gamma^{\nu}$, i.e., $G^{-1}_0(\bm{k},E)=(E+i0)\Gamma^0-\sum^4_{\mu=1}d_{\mu}(\bm{k})\Gamma^{\mu}$. Since the free Green function includes the identity matrix and the Gamma matrices only, we can write the self-energy $\Sigma(E)$ in terms of the identity and Gamma matrices as well, namely, $\Sigma(E)=\Sigma_{0}\Gamma^{0}+\sum^4_{\mu=1}\Sigma_{\mu}\Gamma^{\mu}$. Therefore, 
\begin{equation}
\begin{gathered}
\Sigma(E)=\dfrac{W^2}{12N}\sum_{\bm{q}}\dfrac{1}{(E+i0-\Sigma_0)\Gamma_0-\sum^4_{\mu=1}(d_\mu(\bm{q})+\Sigma_\mu(E))\Gamma^\nu}.
\end{gathered}\label{eq2_13}
\end{equation} 
By comparing the coefficients of the identity and Gamma matrices, we find that
\begin{equation}
\begin{gathered}
\Sigma_0(E)=\dfrac{W^2}{12N}\sum_{\bm{q}}\dfrac{\Sigma_0-E}{\sum^4_{\mu=1}(d_\mu(\bm{q})+\Sigma_\mu(E))^2-(E-\Sigma_0)^2}
\end{gathered}\label{eq2_14}
\end{equation}
and
\begin{equation}
\begin{gathered}
\Sigma_\mu(E)=\dfrac{W^2}{12N}\sum_{\bm{q}}\dfrac{-(d_{\mu}(\bm{q})+\Sigma_\mu(E))}{\sum^4_{\mu=1}(d_\mu(\bm{q})+\Sigma_\mu(E))^2-(E-\Sigma_0)^2}.
\end{gathered}\label{eq2_15}
\end{equation}
For $E=0$, $\Sigma_0$ should be a pure imaginary number, i.e., $\Sigma_0=i(1/\tau)$, from Eq.~\eqref{eq2_14}. Thus, we 
obtain
\begin{equation}
\begin{gathered}
\dfrac{1}{\tau}=\dfrac{1}{\tau}\dfrac{W^2}{12N}\sum_{\bm{q}}\dfrac{1}
{\sum^4_{\mu=1}(d_\mu(\bm{q})+\Sigma_\mu(0))^2+(1/\tau)^2}.
\end{gathered}\label{eq2_16}
\end{equation}
From Eq.~\eqref{eq2_16}, we can determine the critical disorder  $W_{c2}$ for the transition from 3DFOTIs ($1/\tau=0$) to 
DMs ($1/\tau\neq 0$):
\begin{equation}
\begin{gathered}
1=\dfrac{W^2_{c2}}{12N}\sum_{\bm{q}}\dfrac{1}{\sum^4_{\mu=1}(d_\mu(\bm{q})+\Sigma_\mu(0))^2}.
\end{gathered}\label{eq2_16_1}
\end{equation}
Also, the parameter $M$ is renormalized as $\tilde{M}=M+\Delta$ with
\begin{equation}
\begin{gathered}
\Delta=-\dfrac{W^2}{12N}\sum_{\bm{q}}\dfrac{d_2(\bm{q})+\Delta}{\sum^4_{\mu=1}
(d_\mu(\bm{q})+\Sigma_\mu(0))^2}
\end{gathered}\label{eq2_17}
\end{equation}
for both 3DSOTIs and 3DFOTIs. On the other hand, since $d_{1,3,4}(\bm{q})/(D(\bm{q})^2+(1/\tau)^2)$ are odd functions of $\bm{q}$, the integrals of such functions over the first BZ are zeros. Thus, $\Sigma_{1,3,4}=0$.
\par\quad

We use the following scenario to determine $W_{c2}$ numerically. First, we obtain a numerical solution of Eq.~\eqref{eq2_17} by the iterative algorithm:
\begin{enumerate}
\item For a given $M$, choose a fixed disorder $W$.
\item At the first run, set $\Delta_1$ by a random seed value. For the following, we use $\Delta_{i-1}$ of the previous run as the seed.
\item Calculate the right-hand-side (r.h.s) of Eq.~\eqref{eq2_17} by $\Delta_{i-1}$ and set it to be $\Delta_i$.
\item Repeat Steps 2 and 3 for some iterations (in general 50) until $|\Delta_{i}-\Delta_{i-1}|/|\Delta_i|<\delta_{\text{tol}}$ with $\delta_{\text{tol}}$ being the tolerance.
\end{enumerate}
After determining $\Delta$ by the above algorithm, we then calculate the critical disorder $W_{c2}$ for the quantum phase transition from a 3DFOTI to a DM by Eq.~\eqref{eq2_16_1}.
\quad\par

\section{Finite-size scaling analysis}
\label{app3}

The PR of different sizes follow the one-parameter scaling function
\begin{equation}
\begin{gathered}
p_2(L,W)=L^D[f(L/\xi)+CL^{-y}]
\end{gathered}\label{eq3_1}
\end{equation} 
with $D$ being the fractal dimension, $\xi=\xi(W)$ being the correlation length, $y$ being the exponents of the irrelevant
scaling variable, and $C$ being a constant, providing that $W$ is closed to the critical disorder $W_{c3}$ from DMs to
AIs. $f(x)$ is an unknown scaling function, and $\xi$ diverges as a power law near $W_{c3}$, i.e., $\xi\propto|W-W_{c3}|^{-\nu}$ 
with $\nu$ being the critical exponent. 
\par

To obtain the unknown scaling function $f(x)$, we have expanded it to the forth order of $x=L|W-W_{c3}|^{\nu}$ 
\begin{equation}
\begin{gathered}
f(x)=F_0+F_1(L|W-W_{c3}|^{\nu})+F_2(L|W-W_{c3}|^{\nu})^2 \\
+F_3(L|W-W_{c3}|^{\nu})^3+F_4(L|W-W_{c3}|^{\nu})^4
\end{gathered}\label{eq3_2}
\end{equation}
and fitted the numerical data shown in Fig.~4(b) by minimizing the chi square
\begin{equation}
\begin{gathered}
\chi ^2=\sum^{N_w}_{i=1}\sum^{N_L}_{j=1}\left(
\dfrac{p_2(W_i,L_j)-L^D_j[f(L_j|W_i-W_{c3}|^{\nu})+CL^{-y}_j]}{\sigma_{ij}}
\right)^2,
\end{gathered}\label{eq3_3}
\end{equation}
where $N_w$ and $N_L$ are the numbers of disorder strengths and lengths, respectively. $\sigma_{ij}$ is the standard 
deviation of $p_2(W_i,L_j)$. The fitting parameters are $W_{c3},D,\nu,C,y,F_{0,1,2,3,4}$. The fitting yields 
$W_{c3}=18.73\pm0.03$, $D=1.7\pm0.2$, $\nu=1.45\pm0.05$, $C=0.2\pm0.1$, $y=0.7\pm0.1$, and the 
scaling function $f(x)$ defined by Eq.~\eqref{eq3_2}. The goodness-of-fit is $Q=0.2>10^{-3}$, indicating the fit
is acceptable \cite{numRecipes}.

\section{Clean WSM phase}
\label{app4}

\begin{table*}
\caption{\label{tab_gap} This table shows the regimes that Eq.~\eqref{eq2} is gapless, as well as the band touching 
points. Here we fix $B=0.2$.
}
\begin{ruledtabular}
\begin{tabular}{ccc}
Parameter regimes & band-closing points $\bm{k}_0$ & Number of band-closing points in the first BZ \\
\hline
& &  \\
$M\in[3-B,3+B]$ &  $\left(\pm\arccos\left(\dfrac{(M-2)^2+1-B^2}{2(M-2)}\right),0,0\right)$ & 2\\
& &  \\
\hline
& &  \\
\multirow{2}{*}{$M\in[1-B,1+B]$} & $\left(\pm\arccos \left(\dfrac{M^2+1-B^2}{2M}\right),\pm\pi,0\right),
\left(\pm\arccos \left(\dfrac{M^2+1-B^2}{2M}\right),0,\pm\pi
\right)$ & \multirow{2}{*}{6} \\
& $\left(\pm\arccos\left(\dfrac{(M-2)^2+1-B^2}{2(M-2)}\right),0,0
\right)$ & \\
& &  \\
\hline
& &  \\
\multirow{2}{*}{$M\in[-1-B,-1+B]$} & $\left(\pm\arccos \left(\dfrac{M^2+1-B^2}{2M}\right),\pm\pi,0\right),
\left(\pm\arccos \left(\dfrac{M^2+1-B^2}{2M}\right),0,\pm\pi
\right)$ & \multirow{2}{*}{6} \\
& $\left(\pm\arccos \left(\dfrac{(M+2)^2+1-B^2}{2(M+2)}\right),
\pm\pi,\pm\pi\right)$ & \\
& &  \\
\hline
& &  \\
$M\in[-3-B,-3+B]$ &  $\left(\pm\arccos \left(\dfrac{(M+2)^2+1-B^2}
{2(M+2)}\right),\pm\pi,\pm\pi\right)$ & 2 \\
& &  \\
\end{tabular}
\end{ruledtabular}
\end{table*}

In this section, we substantiate that Eq.~\eqref{eq2} is a WSM if $M\in[1-B,1+B]$ and $M\in[3-B,3+B]$ 
by showing that the conduction and the valence bands cross linearly at Weyl nodes in the first BZ. 
The energy spectrum of Eq.~\eqref{eq2} reads
\begin{equation}
\begin{gathered}
E_{pq}(\bm{k})=p\sqrt{\left(B+q\sqrt{d^2_2(\bm{k})+d^2_4(\bm{k})}\right)^2
+d^2_1(\bm{k})+d^2_3(\bm{k})}.
\end{gathered}\label{eq1_10}
\end{equation}
Here $p,q=\pm$ stands for different subbands. To close the gap, one has 
\begin{equation}
\begin{gathered}
B-\sqrt{\left(M-\sum_{i=1,2,3}\cos k_i\right)^2+\sin^2 k_1}=0,
\sin k_2 =0,
\sin k_3 =0.
\end{gathered}\label{eq_w2}
\end{equation}
Note that $\sin k_2=\sin k_3=0$ give $k_2=0,\pm\pi$ and $k_3=0,\pm\pi$. 
We would like to find the possible solutions of Eq.~\eqref{eq_w2} by 
considering different situations. 
\begin{itemize}
\item $(k_2,k_3)=(0,0)$.\par
In this cases, possible $k_1$ and $M$ satisfy  
$B^2=(M-2-\cos k_1)^2+\sin^2k_1$, i.e.,
\[
-1\leq \cos k_1=\dfrac{(M-2)^2+1-B^2}{2(M-2)}\leq 1.
\]
If $(M-2)>0$, one has 
\[
\dfrac{(M-2)^2+1-B^2}{2(M-2)}\leq 1 \to 3-B\leq M \leq 3+B.
\]
Therefore, the gap will close at 
\begin{equation}
\begin{gathered}
\bm{k}_0=\left(\pm\arccos\left(\dfrac{(M-2)^2+1-B^2}{2(M-2)}\right),0,0\right)
\end{gathered}\label{eq_w3}
\end{equation}
for $3-B\leq M\leq 3+B$. If $M=2$, there is no possible $k_1$ and $M$ since $B=0.2\neq 1$. 
If $(M-2)<0$, one has   
\[
-1\leq\dfrac{(M-2)^2+1-B^2}{2(M-2)} \to 1-B\leq M\leq 1+B
\]
Thus, the gap will close at
\begin{equation}
\begin{gathered}
\bm{k}_0=\left(\pm\arccos\left(\dfrac{(M-2)^2+1-B^2}{2(M-2)}\right),0,0\right)
\end{gathered}\label{eq_w4}
\end{equation}
for $1-B\leq M\leq 1+B$. 
\item $(k_2,k_3)=(\pi,0)$ (equivalent to $(-\pi,0),(0,\pi),(0,-\pi)$).\par
Now, we need to find possible $k_1$ and $M$ satisfying 
$B^2=(M-\cos k_1)^2+\sin^2 k_1$, i.e.,
\[
-1\leq\cos k_1 =\dfrac{M^2+1-B^2}{2M}\leq 1.
\]
For $M>0$, 
\[
\dfrac{M^2+1-B^2}{2M} \leq 1\to 1-B\leq M\leq 1+B.
\]
Therefore, the gap will close if $1-B\leq M\leq 1+B$ at
\begin{equation}
\begin{gathered}
\bm{k}_0=
\left(\pm\arccos \left(\dfrac{M^2+1-B^2}{2M}\right),\pm\pi,0\right),\\
\left(\pm\arccos \left(\dfrac{M^2+1-B^2}{2M}\right),0,\pm\pi\right).
\end{gathered}\label{eq_w5}
\end{equation}
For $M=0$, no $k_1$ and $M$ exist since $B\neq 1$. For $M<0$, 
\[
-1\leq\dfrac{M^2+1-B^2}{2M} \to -1-B\leq M\leq -1+B.
\]
In this cases, the gap will close if $-1-B\leq M \leq -1+B$ at
\begin{equation}
\begin{gathered}
\bm{k}_0=\left(\pm\arccos \left(\dfrac{M^2+1-B^2}{2M}\right),\pm\pi,0\right),\\
\left(\pm\arccos \left(\dfrac{M^2+1-B^2}{2M}\right),0,\pm\pi\right).
\end{gathered}\label{eq_w6}
\end{equation}
\item $(k_2,k_3)=(\pi,\pi)$ (same as $(\pi,-\pi),(-\pi,\pi),(-\pi,-\pi)$).\par
Now, it is required that $B^2=(M+2-\cos k_1)^2+\sin^2 k_1$, i.e.,
\[
-1\leq\cos k_1=\dfrac{(M+2)^2+1-B^2}{2(M+2)}\leq 1.
\]
If $M+2>0$, 
\[
\dfrac{(M+2)^2+1-B^2}{2(M+2)}\leq 1 \to -1-B\leq M\leq -1+B.
\]
Therefore, the gap closes for $-1-B\leq M\leq -1+B$ at
\begin{equation}
\begin{gathered}
\bm{k}_0=\left(\pm\arccos \left(\dfrac{(M+2)^2+1-B^2}{2(M+2)}\right),\pm\pi,\pm\pi\right).
\end{gathered}\label{eq_w7}
\end{equation}
For $M+2=0$, the gap is open since $B=0.2\neq 1$. For $M+2<0$,
\[
-1\leq \dfrac{(M+2)^2+1-B^2}{2(M+2)} \to -3-B\leq M\leq -3+B.
\]
Therefore, the gap closes for $-3-B\leq M\leq -3+B$ at 
\begin{equation}
\begin{gathered}
\bm{k}_0=\left(\pm\arccos \left(\dfrac{(M+2)^2+1-B^2}{2(M+2)}\right),\pm\pi,\pm\pi\right).
\end{gathered}\label{eq_w8}
\end{equation}
\end{itemize}
We summarize all those results in Table~\ref{tab_gap}.
\par

We then discuss the dispersion near the band touching points $\bm{k}_0$ by expanding the dispersion 
Eq.~\eqref{eq1_10} about $\bm{k}=\bm{k}_0+\delta\bm{k}$. Let us consider the parameter regime 
$M\in[3-B,3+B]$ where the bulk touching points are (refer Table~\ref{tab_gap})
\begin{equation}
\begin{gathered}
\bm{k}_0=\left(\pm\arccos\left(\dfrac{(M-2)^2+1-B^2}{2(M-2)}\right),0,0\right)\\
=\left(\pm\arccos k_{0,1},0,0\right).
\end{gathered}\label{eq_w9}
\end{equation}
The conduction and valence bands read
\begin{equation}
\begin{gathered}
E_{p-}(\delta\bm{k})=p\sqrt{\left(\dfrac{(M-2)\sin k_{0,1}}{B}\right)^2(\delta k_1)^2+(\delta k_2)^2+(\delta k_3)^2}.
\end{gathered}\label{eq_w10}
\end{equation}
which cross linearly at the band touch points. Likewise, we obtain the same dispersion as Eq.~\eqref{eq_w10} for 
$\delta \bm{k}=\bm{k}-(-k_{0,1},0,0)$. Therefore, for $B\neq 0$ and $M\in[3-B,3+B]$, the system is a WSM 
with linear dispersions near $(\pm k_{0,1},0,0)$. Following the same approach, we find the system to be a WSM 
for $-3-B\leq M\leq -3+B$ (with a pair of Weyl nodes), as well as $-1-B\leq M\leq -1+B$ and $1-B\leq M\leq 1+B$ 
(with three pairs of Weyl nodes). 
\par

\end{document}